\documentclass{aa}
\usepackage{psfig,times}
\pssilent
\begin{document}
\thesaurus{section number (02.18.5; 02.18.6; 08.16.7 Vela)}
\title{Post-glitch RXTE-PCA observations of the Vela pulsar}
\author{M. Atakan G\"urkan\inst{1}\thanks{\emph{Present address:} Northwestern 
University, Department of Physics and Astronomy, 2145 Sheridan Road, Evanston, IL 60208, USA.} 
  \and Altan Baykal\inst{1}
  \and M. Ali Alpar\inst{1}\thanks{\emph{Present address:} Faculty of Sciences 
and Engineering, Sabanc\i\ University, Karak\"oy, \.Istanbul, Turkey. }  
  \and Hakk\i\ B. \"Ogelman\inst{2}
  \and Tod Strohmayer\inst{3} }
\institute{Middle East Technical University, Physics Department, 06531, Ankara, Turkey.
  \and Department of Physics, University of Wisconsin, 1150 University Ave., Madison, 
  WI 53706, USA 
  \and NASA, Goddard Space Flight Center, Greenbelt, MD 20771, USA. }
\offprints{A. G\"urkan}
\mail{ato@nwu.edu}
\date{Received 18 Aug 1999 / Accepted 25 Feb 2000}
\maketitle
\begin{abstract}
We report the results of analysis of observations of the 
Vela Pulsar by PCA on RXTE.
Our data consists of two parts. The first part contains observations at 1, 4,
and 9 days  after the glitch in 
1996 and has 27000 sec.  total 
exposure time. The second part of observations were
performed 
three months after this glitch and have a total exposure time of 93000 sec. 
We found pulsations in both sets. The observed spectrum is a power-law with
no apparent change in flux or count rate. The theoretical expectations 
of increase in flux due to internal heating after a glitch are smaller 
than the uncertainty of the observations. 
\end{abstract}
\section{Introduction}
\label{velaintro}
We present observations of the \object{Vela pulsar} with the Proportional
Counter Array (PCA), on the Rossi X-Ray Timing Explorer (RXTE). Our observations
cover two distinct time-spans. The first part is very close to the glitch on 
1996 October 13.394 UT (Flanagan \cite{glidate}). It consists of three 
observations at one, four and nine days after the glitch. We analyzed
these sets of data separately. The second series of observations were
obtained in January 1997. All data sets of January 1997 were analyzed together. 
The exact dates of observations are given in Table~\ref{the_table}.

We first performed spectral analysis of our data,
calculated time averaged flux for different observations and
put upper limits for the flux change. Then, by using radio ephemerides, we
detected the pulsations in the data, and investigated the changes
in pulse shape and pulse fraction. Finally, we compared our results with the 
theoretical expectations of change in flux which might arise because of glitch
induced energy dissipation in the neutron star. 

Time averaged spectrum analysis is explained in section \ref{pavspec}. The
detected pulse shapes are presented in section 
\ref{velatiming}. 
In section \ref{velaconc} we discuss the implications of our
results.
\section{Time Averaged Spectrum Analysis}
\label{pavspec}
The observation time-spans, and total integration times are gi\-ven in 
Table~\ref{the_table} together with calculated model parameters, and flux
values.
The analysis is carried out using FTOOLS~4.1.1 and XSPEC~v10.
Only the data coming from the 
first xenon layer was chosen to increase the signal-to-noise
ratio. The time intervals in which one or more of the five Proportional
Counting Units (PCUs) are off, the elevation angle is less than 
10 degrees, or pointing 
offset is greater than 0.02 degrees were not included in the analysis, 
as recommended in the ``Screening'' section of ``ABC of XTE'' 
(RXTE GOF \cite{abcxte}). 
The background used is synthetic and is generated by the background estimator
{\tt pcabackest}. The background models are 
based on rate of very large events, spacecraft activation, and cosmic
X-ray emission. 
More information on background models can be
found in (Jahoda \cite{jaho}). We have used the 2.2.1\_v80 version
of response matrices. Although the matrices are not equally good for each PCU,
to have good statistics, we combined data coming from every PCU, contrary to the
recommendation by Remillard (\cite{remi_resp}).

\begin{figure}
\psfig{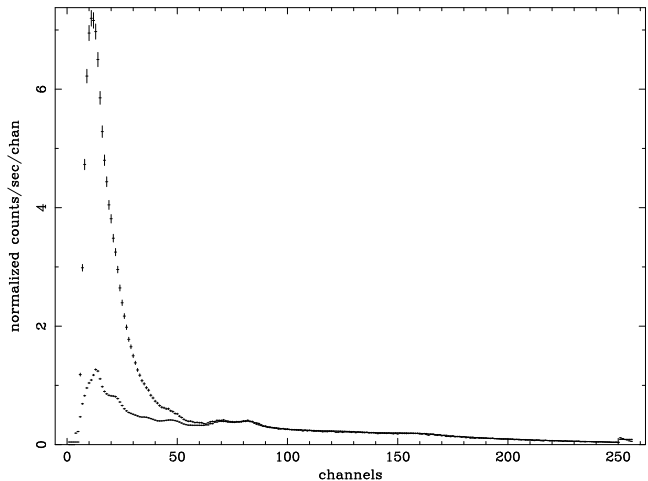}
\caption{Number of counts versus channel for the data and the background. 
After around channel 68 the data and the background coincide.\label{datbac}}
\end{figure}

A comparison of background with the data led us to ignore the channels above
68, which approximately corresponds to the energy
25.7 keV (see Fig.~\ref{datbac}.). 
The systematic errors were chosen to make the reduced $\chi^2$ equal to
unity in XSPEC. To have reasonable systematic errors we also had to ignore
channels 0-7. The maximum energy for the seventh channel is 2.90 keV.

The hydrogen column density model used by XSPEC is valid for the 
energies 0.03-10 keV. Although this covers the ROSAT energy band (0.5-2.4 keV),
major portion of our spectrum (2.90-25.7 keV) falls outside this range.
Therefore, we adopted the hydrogen column density $4\times
10^{20}\hbox{atoms}/\hbox{cm}^2$ obtained in a ROSAT observation of the Vela
pulsar (\"Ogelman et al. \cite{vel_og}), since at lower energies the spectral
resolution of ROSAT is much better than RXTE/PCA detectors' resolution.

\begin{figure*}
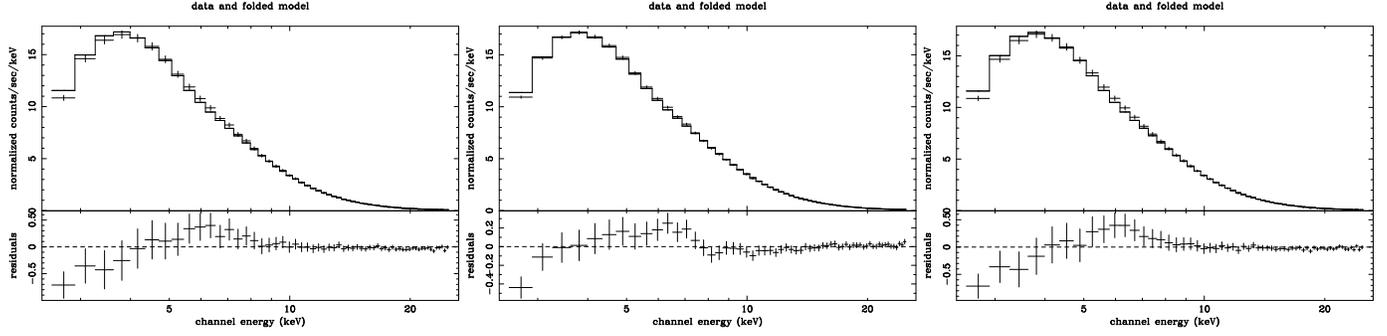

\hbox{
\psfig{figure=fig2-1.ps,width=6cm,angle=270}
\psfig{figure=fig2-2.ps,width=6cm,angle=270}
\psfig{figure=fig2-3.ps,width=6cm,angle=270}}
\caption{Spectra of observations 10276-01-01-00, 10276-01-02-00, and 
10276-01-03-00. The horizontal axis is energy
in keV, the vertical axis shows counts/sec/keV (upper panel) and calculated
residuals (lower panel). The solid line 
denotes the fitted model
(photon absorbed power-law).\label{spec1}}
\end{figure*}

\begin{figure*}
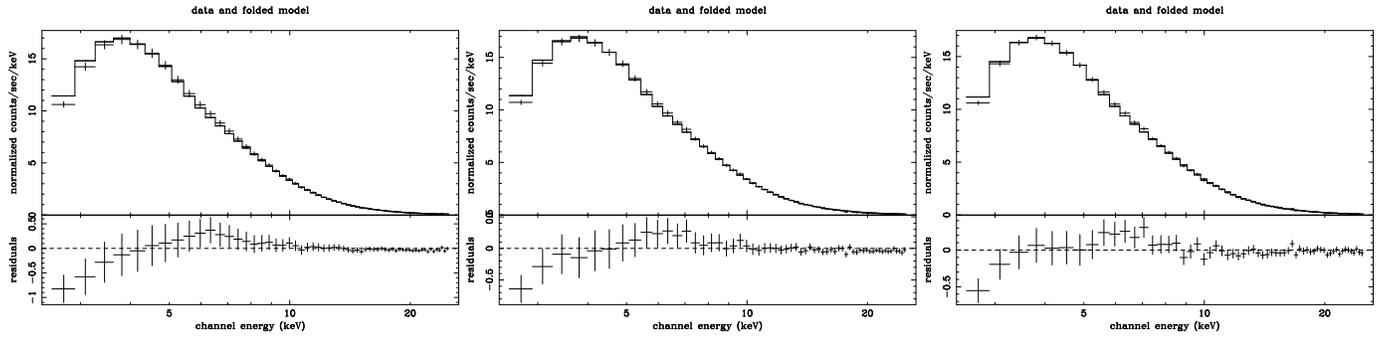

\hbox{
\psfig{figure=fig3-1.ps,width=6cm,angle=270}
\psfig{figure=fig3-2.ps,width=6cm,angle=270}
\psfig{figure=fig3-3.ps,width=6cm,angle=270}}
\caption{Spectra of observations 10275-01-01-00,  
10275-01-01-03, and 10275-01-01-06. 
The horizontal axis is energy
in keV, the vertical axis shows counts/sec/keV (upper panel) and calculated
residuals (lower panel). The solid line 
denotes the fitted model
(photon absorbed power-law). The results obtained from other observations in
January 1997 are very similar to these presented results.\label{spec2}}
\end{figure*}

Figures \ref{spec1} and \ref{spec2} are plots of the energy spectrum along
with fitted models and residuals. The model parameters are given in 
Table~\ref{the_table}. The quoted errors are for three sigma confidence levels. 
The calculated
flux for each observation along with an upper and a 
lower limit are given in Table~\ref{the_table}. 
The upper (lower) limits for the flux 
are calculated by setting the index of the power-law to the 
lower (upper) limits
given by XSPEC, leaving the normalization of the power-law spectrum
as the only free parameter, and refitting the spectrum.

\begin{table*}
\caption{Parameters for the observations. The third column is the total time
devoted to the observation in seconds. The power-law index values have three 
$\sigma$ confidence levels.
The unit for flux is 
$10^{-10}\hbox{ergs}\,\hbox{cm}^{-2}\hbox{sec}^{-1}$. The last column is the
percent systematic error.\label{the_table}}
\begin{center}
\begin{tabular}{llllllll}
\hline
Observation ID & Date & time & power-law &\multicolumn{3}{c}{Flux for 2-20 keV} & sys. \\ 
 & & (secs) & index & Low Lim. & Value & Up Lim. & err.\\ \hline
10276-01-01-00 & 14/10/96 &  8434 & $2.073^{2.101}_{2.045}$ & 2.735 & 2.741 & 2.747 & 1.8\\
10276-01-02-00 & 17/10/96 &  8762 & $2.024^{2.038}_{2.008}$ & 2.801 & 2.805 & 2.804 & 0.7\\
10276-01-03-00 & 22/10/96 &  9601 & $2.067^{2.092}_{2.044}$ & 2.758 & 2.763 & 2.768 & 1.6\\
10275-01-01-00 & 17/01/97 & 17707 & $2.077^{2.107}_{2.046}$ & 2.697 & 2.707 & 2.716 & 2.1\\
10275-01-01-01 & 22/01/97 & 14808 & $2.066^{2.091}_{2.039}$ & 2.712 & 2.719 & 2.725 & 1.7\\
10275-01-01-02 & 17/01/97 &  5363 &  $2.071^{2.100}_{2.041}$& 2.713 & 2.718 & 2.722 & 1.8\\
10275-01-01-03 & 18/01/97 &  6601 &  $2.060^{2.086}_{2.033}$& 2.722 & 2.724 & 2.727 & 1.5\\
10275-01-01-04 & 23/01/97 &  8411 &  $2.059^{2.083}_{2.034}$& 2.715 & 2.719 & 2.722 & 1.5\\
10275-01-01-05 & 13/01/97 &  7228 &  $2.045^{2.064}_{2.025}$& 2.739 & 2.739 & 2.740 & 1.0\\
10275-01-01-06 & 12/01/97 &  4379 &  $2.044^{2.065}_{2.022}$& 2.708 & 2.711 & 2.713 & 1.0\\
10275-01-01-07 & 20/01/97 & 11695 &  $2.062^{2.085}_{2.037}$& 2.710 & 2.714 & 2.719 & 1.5\\
10275-01-01-08 \& 080 & 21/01/97 & 16765 & $2.070^{2.095}_{2.044}$   & 2.716 & 2.723 & 2.730&  1.7\\
\hline
\end{tabular}
\end{center}
\end{table*}
We have also searched for a blackbody component to the spectrum in addition to
power-law, but this resulted in temperatures 
$\sim\!\!10$ times larger than what has
been found by \"Ogelman et al. (\cite{vel_og}), for {\em all\/} 
observations. Although the addition of a blackbody component improves the fit,
the resulting temperature suggests that this is not physical but merely a result
of an increase in the number of variables. This idea is supported by the fact
that adding a bremsstrahlung component to power-law or changing the power-law
to a broken power-law gives a fit as good as a power-law blackbody combination.
\section{Timing Analysis}
\label{velatiming}
The data analyzed consists of two parts.
The earlier data sets extend between one and ten days after the glitch, where 
the post-glitch exponential relaxation of the pulse period prevails. 
The second data set, about three months after the glitch, does not display
this rapid variation of the pulse frequency.

\begin{figure*}
\psfig{figure=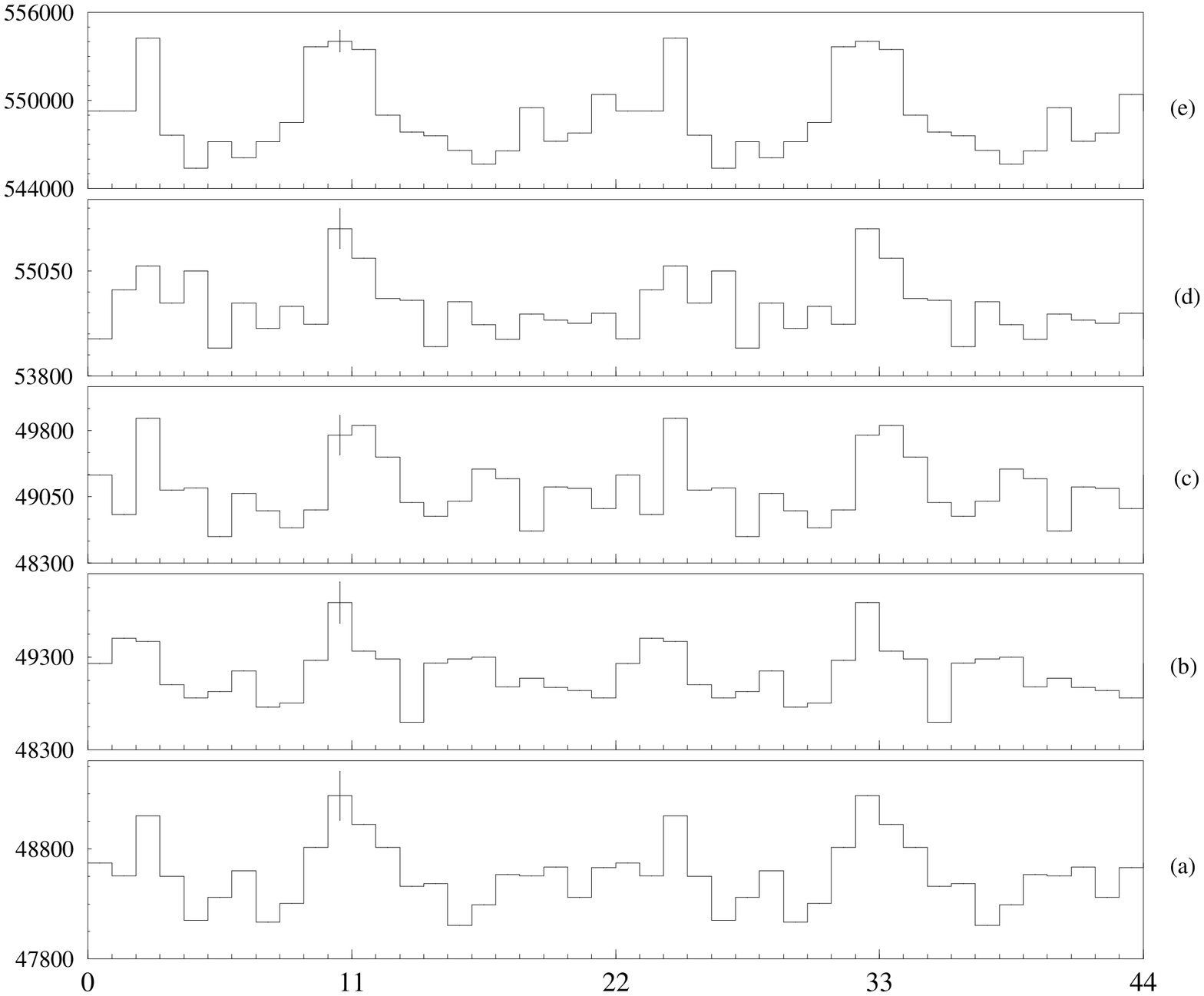,angle=270,width=\hsize}
\caption{Pulse shapes obtained for the observations. 
$y$-axis is number of photon counts, $x$-axis is twice 
the phase. The photon counts
include {\em all\/} photons detected in the first layer, channels 8-68
inclusive (2.90-25.7 keV). 
No background subtraction was done. A single, typical error bar is
displayed for each set.
({\bf a}) and ({\bf b}) are the first
two observations just after the glitch, these curves are obtained by 
interpolating values of frequency and frequency derivative given by the 
ephemerides. ({\bf c}) and ({\bf d}) are the second and third
observations, these curves are obtained by polynomial fit to values given in
the Princeton ephemerides. ({\bf e}) represents all the observations 
three months after the glitch, this curve is obtained by using the values given
in the Princeton ephemerides.
\label{pulses}}
\end{figure*}

Finding a pulsation in the latter was straightforward. By using the Princeton
ephemerides distributed with FTOOLS the pulse shape shown in  
Fig.~\ref{pulses}(e) is obtained. This is a histogram of 
counts versus twice the
phase, which is divided into 22 intervals.  
The histogram includes all
photons detected in the first layer and in channels 8-68 inclusive.
No filtering was done for elevation, offset or number of PCUs. Since background 
is synthetic it is not subtracted either.

\begin{figure}
\psfig{figure=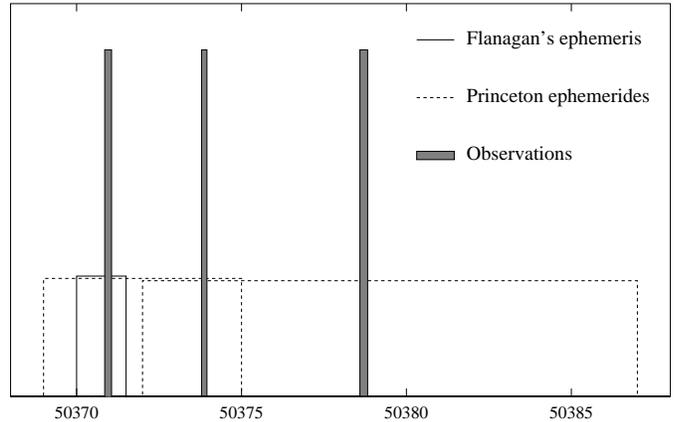,angle=0,width=88mm}
\caption{Observation and ephemerides time-spans. The rectangular spikes
match the duration of the observations. 
Solid line is ephemeris provided by Flanagan.
Dashed lines denote Princeton ephemerides. The horizontal axis is marked in
Modified Julian Day.
\label{obseph}}
\end{figure}

Finding a pulse for earlier observations, which are very close to the 
glitch, proved to be difficult. There are two sets of ephemerides in the
Princeton database that are relevant to these observations, 
and an additional one was provided by Claire Flanagan (private communication). 
The first two give the frequency, frequency derivative, and
frequency second derivative, while the third one gives only frequency and 
frequency derivative. The time-spans covered by these 
ephemerides and the three observations in the earlier set 
are shown in Fig.~\ref{obseph}. The reference epoch of the
ephemerides are roughly 50370.7, 50372.0, and 50379.0
for Flanagan and Princeton ephemerides one and two, respectively.
None of the ephemerides by itself gives a pulse for any of the three 
observations. 
Furthermore the ephemerides give different results for overlapping portions.

We therefore tried to combine the ephemerides. We interpolated the values
of frequency and frequency derivative by making a second order polynomial
fit to the values given in the ephemerides. This yielded to the expressions:
\begin{eqnarray}
f &=& .2975280743\times10^{-18}t^2 \nonumber \\
& &-.16020626525\times 10^{-10}t+11.1962177095, \label{puleq1}\\
\dot f &=& -.242915\times 10^{-23}t^2 \nonumber \\
& &+.242121\times 10^{-17}t
-.1622\times 10^{-10}\!\!. \label{puleq2}
\end{eqnarray}
The epoch for these expressions is the same Flanagan's ephemeris.
By using these values and not taking the higher derivatives into account, we
calculated the phase for the arrival time of each photon. This method yielded
reasonable pulse shapes for the first and second observation (see 
Fig.~\ref{pulses} (a) and (b) ), but failed for
the third one. The reference epoch of the second Princeton ephemeris is very
near to the third observation, nevertheless the use of the ephemeris which 
represents an extended time-span of rapidly varying periods, fails to give a
pulse shape by itself. 
We therefore tried another
approach. We combined the frequencies, frequency derivatives, 
and frequency second
derivatives given in the two Princeton ephemerides, made a fifth order
polynomial fit, 
and threw away the fourth and fifth order terms. In this way we reduced
the contribution of the second ephemeris. The final expression for frequency
is:
\begin{eqnarray}
\label{jfereq}
f &=& -9.68274932\times 10^{-25} t^3 + 0.8\times 10^{-18} t^2 \nonumber \\
& & - 1.59821\times 10^{-11}t +  11.1962159427143.
\end{eqnarray} 
The epoch for this expression is the same as first Princeton ephemeris.
The pulse shapes obtained for the
second and third observations by this method are given in Fig.~\ref{pulses}~(c)
and (d). 

\section{Conclusions and Discussion}
\label{velaconc}
\subsection{Time Averaged Spectrum}
The power-law spectrum observed is in agreement 
with expectations deduced from previous 
observations of Vela at higher and lower 
energies(\"Ogelman et al. \cite{vel_og}, Kanbach et al. \cite{kanbach},
Strickman et al. \cite{strick96}, 
Kuiper et al. \cite{kuiper} ). At this part of spectrum (2-20 keV), the
contribution of the pulsar is 
very small compared to the contribution of the compact nebula
surrounding it. As a result the pulse shapes have 
a very high DC level, as can be seen in
Fig.~\ref{pulses}.

The slightly higher residuals near 6 keV and lower 
residuals near 4 keV are not characteristics
of observed sources, but are artifacts of PCA. 
This effect, which is a result of 
the L edge of Xenon, is reduced by 
the version of 
response matrices in use, but not completely removed.

Our main conclusion from the analysis is that the spectrum does not
change from early post-glitch to late observations. 
It is a power-law with an index around 2 for all of the observations.
The power-law index does not change significantly among the 
observations. The highest value calculated for the index is 2.107 and 
the lowest value is 2.009.  This corresponds to a change of 5\%,
which is a fractional upper limit
for the change of power-index during the observations.

The upper and lower limits of the flux calculated by the comparison
explained in section \ref{pavspec} and presented in Table~\ref{the_table}
are well 
within the range of systematic errors. We therefore adopt the systematic errors
as the upper limits to any variation in flux. 

There is seemingly a jump in the 
flux between 2-20 keV, from the first to the 
second observation. This observation 
is only four days away from the glitch.
The pre-glitch temperature of the surface of the Vela pulsar is thought to be
around 0.15 keV(\"Ogelman et al. \cite{vel_og}). Theoretical models
(Van Riper et al.\cite{nsth1}; Umeda et al.\cite{nsth2}; Hirano et al.
\cite{nsth3}) predict an increase at most by a factor
of 8, which brings the temperature to 1.2 keV. Attempts to find a blackbody
component in this observation did not give significantly different results
from other observations. This suggests that the observed 
flux changes may have little or nothing
to do with changes in surface temperature. 

Another possible interpretation
is that there is an error in the analysis of this 
particular observation, possibly arising from the
calculation of synthetic background. Vela is a faint source for 
PCA. An improved model in the estimation of
background for faint sources has been released by the PCA Team in 1998. 
This model
has been used throughout the calculations. There may be further improvements
on the background models that could change the calculated flux. The presented
flux is calculated by using the spectrum model, rather than by direct
observation. 
Apart from this observation there is no apparent change
in flux or count rate.

Treating the calculated fluxes as very high upper bounds to
the Wien tail of possible blackbody radiation from the neutron star surface
could in principle be used to rule out some of the models for the
post-glitch thermal emission from the neutron stars. In practice this does
not work since the surface 
temperature range of the Vela pulsar is far below the
RXTE-PCA energies.
\subsection{Timing Analysis and Pulse Shapes}
The epoch of the second ephemeris taken from the Princeton database, 
50379.0 MJD, is pretty close
to the third observation (see Fig.~\ref{obseph}), but using the ephemeris
alone for the observation does not give a pulse shape. This may be due to
the existence of two distinct decay time scales of Vela, 3 days and 30 days, 
which were observed
in all previous glitches and fall within the ranges of ephemeris (Alpar et al. 
\cite{Alpar93}). Our data is not good enough to determine any exponential
decay time scales.

In view of the rapidly varying period at those epochs,
the pulse shapes of the first part of the observations were obtained by a
careful interpolation amounting to the construction of an ephemeris
that can represent the rapid changes in the pulsar's timing parameters
in this postglitch epoch. The phase difference of the two observed peaks
in these shapes is the same as the phase difference in 
the second set of observations (in January 1997). This gives us some
confidence in the resultant pulse shapes.

The pulse shapes obtained are not reliable for drawing conclusions 
on the changes of pulse shape or
pulsed fraction, since both of these factors are sensitively dependent on
ephemeris. This is best seen by comparing Fig.~\ref{pulses}~(b) 
and (c). 
They belong to the same set of data but have obvious differences 
both in the pulse shape and pulsed fraction.

\subsection{Possible Future Work}
Extracting the contribution of the compact nebula from the spectrum may help
to delineate effects of temperature changes on the neutron star
surface. Although the pulsations are detected, they are not reliable enough to 
justify taking the off-peak
photon counts as background to the peak photon counts to remove the effects
coming from the DC signal.
The field of view of the PCA detector is one degree (RXTE GOF \cite{PCA}), 
consequently the compact nebula surrounding the Vela pulsar has a significant
contribution to the observed spectrum. 
The images showing the 
emission from the pulsar and the sources around it, in particular the compact
nebula, can be found in Markwardt (\cite{craig_neb}),
Frail et al. (\cite{ogel2}), Harnden et al. (\cite{1985}), and Willmore et al.
(\cite{willmore}). 

When we divided the data from the second part of observations into smaller time 
intervals we have observed that the pulse shape begins to disappear 
for data strings covering less than 30000 seconds. The exposure time of the
data sets we used in this work are below 10000 seconds. This explains the
uncertainty in pulse shapes and fractions. Future target of opportunity 
observation of the Vela pulsar by RXTE need to be allocated more
observation time, and should contain observations made approximately 
20 days after the
glitch, since this is about the time that the surface temperature will
reach its maximum according to theoretical models. Also, more detailed 
ephemerides fitting the post-glitch behavior of the pulsar is necessary to make
deductions on changes in pulse shape.

Finally we note that the question of glitch associated energy 
dissipation in the Vela pulsar has been addressed also with ROSAT 
observations. Comparison of observations at epochs before and after the 
glitch has not yielded stringent constraints on the glitch related energy 
dissipation (Seward et al. \cite{ROSAT}).

While this work was in preparation another analysis of RXTE/PCA observations
of the Vela pulsar was published by Strickman et al. (\cite{strick2}).
They have also detected a pulsed emission and a power-law spectrum. 
Our analysis differs from theirs in two ways. Their
phase-resolved spectra are obtained by taking ``off-pulse'' 
photons as background
to ``on-pulse'' photons, whereas we calculated only time averaged spectra.
Another difference is that these authors used data coming from only the 
first xenon 
layer for energies below 8 keV, but included data coming from the other two
layers for higher energies. In our analysis we used photons detected only in
the first xenon layer. As a result of these differences, their power-law index 
is smaller than the value that we found.
\begin{acknowledgements}
We thank Claire Flanagan for providing the ephemeris, Sally K. Goff for
helping the preparation of the manuscript, and an anonymous referee for
useful comments. Some calculations in this paper
were performed on the ``tasman'' computer at 
METU Computer Center which was made available
by \c Ca\u gr\i\ \c C\"oltekin. This analysis was made possible with
the help and documentation provided by RXTE-PCA team, for which we thank the
members of the team, in particular Keith Jahoda. 
M.A.G., A.B., M.A.A. and H.B.\"O.  
acknowledge support from the Scientific and Technical Research council of 
Turkey, T\"UB\.ITAK, under grant TBAG-\"U 18. M.A.G. also acknowledges a
scholarship provided by T\"UB\.ITAK, and partial support from National Science
Foundation (DMR 91-20000) through the Science and Technology Center for
Superconductivity. M.A.A. also acknowledges 
support from the Turkish Academy of Sciences.
\end{acknowledgements}

\end{document}